\documentclass[twocolumn,prc,aps,floatfix]{revtex4}
\usepackage[dvips]{graphicx}
\begin{document}
\newcommand{\be}{\begin{eqnarray}}
\newcommand{\ee}{\end{eqnarray}}
\title{Enhanced effect of temporal variation of the fine structure constant
and strong interaction in $^{229}$Th}
\author{V.V. Flambaum}
\affiliation{
 School of Physics, The University of New South Wales, Sydney NSW
2052, Australia
}
\date{\today}
\begin{abstract}
The relative effects of variation of the fine structure constant
 $\alpha=e^2/\hbar c$ and dimensionless strong interaction parameter
$m_q/\Lambda_{QCD}$ are enhanced 5-6 orders of magnitude in
 very narrow
 ultraviolet transition between the
ground and first excited state in $^{229}$Th nucleus
(energy $(3.5\pm 1)$ eV). Corresponding experiment has potential
of improving sensitivity to the variation of the fundamental
constants by 7-10 orders of magnitude (up to $10^{-23}$ per year).
\end{abstract}
\maketitle

Unification  theories applied to cosmology suggest a possibility
of variation of the fundamental constants in
the expanding Universe (see e.g. review \cite{Uzan}).
There are  hints of variation of
 $\alpha$ and $m_{q,e}/\Lambda_{QCD}$
in quasar absorption spectra, Big Bang nucleosynthesis and Oklo
natural nuclear reactor data
(see \cite{var} and references therein).
 Here  $\Lambda_{QCD}$ is the quantum chromodynamics
 (QCD) scale,  and $m_{q}$ and $m_{e}$ are the quark and electron masses.
However, the majority of publications report only limits on  possible variations
 (see e.g. reviews \cite{Uzan,karshenboim}).
 A very sensitive method to study the
 variation in a laboratory
 consists of the comparison of different optical and microwave atomic clocks
(see  recent measurements in
 \cite{prestage,Marion 2003,Bize 2005,Peik 2004,Bize 2003,
Fischer 2004,Peik 2005}).
An  enhancement of the relative effect of $\alpha$ variation can be obtained
in a transition between almost degenerate levels in Dy atom \cite{dzuba1999}.
These levels move in opposite directions if  $\alpha$ varies.
 An experiment is currently underway to place limits on
$\alpha$ variation using this transition \cite{budker}, but
unfortunately one of the levels has  quite a large linewidth
and this limits the accuracy. An enhancement of 1-3 orders exists
in narrow microwave molecular transitions \cite{mol}.
Some atomic transitions with enhanced sensitivity are listed in
Ref.~\cite{at}. 

A very  narrow level  $(3.5\pm 1)$ eV above the ground state exists
 in $^{229}$Th nucleus \cite{th} (in \cite{th1} the energy is
 $(3.67\pm 0.03)$ eV, in  \cite{th6} $(5.5\pm 1)$ eV ). The width of this level is estimated to be
about $10^{-4}$ Hz \cite{th2} (the experimental limits on the width are given
in \cite{th3}). This makes $^{229}$Th a possible reference for an
 optical clock of very high accuracy, and opens a new possibility 
for a laboratory search for the varitation of the fundamental constants
\cite{th4}.
 Below I will show that there is an additional very important advantage.
The relative effects of variation of
 $\alpha$ and $m_{q}/\Lambda_{QCD}$ are enhanced by 5-6 orders of magnitude.

  The ground state of  $^{229}$Th nucleus is $J^P[Nn_z\Lambda]=5/2^+[633]$;
i.e. the deformed oscillator quantum numbers are $N=6$, $n_z=3$, the
 projection of the valence neutron orbital angular momentum on the
nuclear symmetry axis (internal  z-axis)
 is $\Lambda=3$, the spin projection $\Sigma=-1/2$, and the
 total angular momentum and the total 
 angular momentum projection are $J=\Omega=\Lambda+\Sigma=5/2$.
The 3.5 eV excited state is $J^P[Nn_z\Lambda]=3/2^+[631]$; i.e.
it has the same  $N=6$ and  $n_z=3$. The values
$\Lambda=1$, $\Sigma=1/2$ and  $J=\Omega=3/2$ are different.
 The energy of both states
may be described by an equation \cite{BM}
$E=E_0 + C \Lambda \Sigma + D \Lambda ^2$,
i.e. the energy difference between the excited and ground state
is $\omega=E_e-E_g=2 C - 8D$. The values of the constants
$C$ and $D$ are presented, for example, in the book \cite{BM}.
Note that  $\omega$ is 5 orders of magnitude smaller than $C$ and $D$.
Therefore, for consistency of this simple valence model,
 we must take  $2 C \approx  8D$.
Based on the data from \cite{BM} we will use the following numbers:
 $2 C \approx  8D \approx -1.4$ MeV. The relative variation of
the transition frequency may be presented as
\begin{equation}\label{delta1}
 \frac{\delta \omega}{\omega}=\frac{\delta (2C)- \delta (8D)}{\omega}
\approx 0.4 \cdot 10^6 (\frac{\delta D}{D}-\frac{\delta C}{C})
\end{equation}  
The large factor here appeared from the  ratio
 $2C/\omega \approx 8D/\omega \approx -0.4 \cdot 10^6$ for $\omega$=3.5 eV.
The orbit-axis interaction constant $D$ vanishes for zero
deformation parameter $\beta_2$.
 Therefore, we should assume that $D \approx const \cdot V_0 \beta_2$
where $V_0$ is the depth of the strong potential.
The nuclear deformation reduces the energy of the Coulomb
repulsion between the protons. Without this repulsion
the deformation parameter $\beta_2$  would  probably 
be zero
\cite{comment}. Therefore, it is natural to assume that
 $\beta_2 \approx const  \cdot \alpha$.
 Thus we have  $D \approx const \cdot V_0 \cdot \alpha$ and   
\begin{equation}\label{deltaD}
\frac{\delta D}{D} \approx \frac{\delta V_0}{V_0} +
\frac{\delta \alpha}{\alpha}
\end{equation}  
To estimate variation of $V_0$ we will use
 Walecka model \cite{SW} where the strong nuclear potential is produced by
 the sigma and the omega meson exchanges 
\begin{equation}\label{walecka}
V= -{g_s^2 \over 4\pi} {e^{-r m_\sigma} \over r}+{g_v^2\over 4\pi}
 {e^{-r m_\omega} \over r}
\end{equation} 
Using eq. (\ref{walecka}) we can find the depth of the potential well
 \cite{FS,comment1}
\begin{equation}\label{depth}
 V_0= \frac{3}{4 \pi r_0^3} \left(\frac{g_s^2}{m_\sigma ^2} -
\frac{g_v^2}{m_\omega ^2} \right)
\end{equation}
Here $r_0$= 1.2 fm is an inter-nucleon distance.
Note that the nuclear potential in this model is  a highly tuned
 small difference of two large terms. 
Therefore, the contribution of the variation of $r_0$
 is not as important as the  contribution of the meson mass variation
which is enhanced due to the cancellation of two terms in $V_0$.
The result is \cite{FS}
\begin{equation}{\delta V_0 \over V_0} \approx - 8.6 \frac{\delta m_\sigma}{m_\sigma}+ 
 6.6 \frac{\delta m_\omega}{m_\omega}
\end{equation}
The final result will depend on the variation of the dimensionless
parameter $m_q/\Lambda_{QCD}$.
During the following calculations, we shall assume
 that  $\Lambda_{QCD}$ does not vary and so
 we shall speak about the variation of masses (this means that we measure
 masses in units of $\Lambda_{QCD}$). We shall restore the explicit
 appearance of $\Lambda_{QCD}$ in the final answers. The dependence
of the meson masses on the  current light quark
  mass $m_q=(m_u+m_d)/2$ has been calculated in  Ref. \cite{roberts}
$\frac{\delta m_{\sigma}}{m_{\sigma}} = 0.013 \frac{\delta m_q}{m_q }$ , 
$\frac{\delta m_{\omega}}{m_{\omega}} = 0.034 \frac{\delta m_q}{m_q }$. 
This gives us
 \begin{equation}{\delta V_0 \over V_0} \approx 0.11 \frac{\delta m_q}{m_q}
\end{equation}
The relatively small contribution of the light quark mass is explained
by the fact that $m_q \approx 5$ MeV is very small. The contribution
of the strange quark mass $m_s \approx$ 120 MeV  may be much larger.
According to the calculation in Ref.~\cite{FS}
${\delta  m_\sigma\over m_\sigma}\approx 
0.54  {\delta m_s \over m_s}$ ,
${\delta  m_\omega\over m_\omega}\approx 0.15  {\delta m_s \over m_s}$ ,  and
${\delta V_0 \over V_0} \approx -3.5 \frac{\delta m_s}{m_s}$.
By adding all contributions we obtain
\begin{equation}\label{deltaD1}
\frac{\delta D}{D} \approx 
\frac{\delta \alpha}{\alpha} + 0.11 \frac{\delta m_q}{m_q }
 -3.5 {\delta m_s \over m_s}
\end{equation}  
The reason for the enhancement here ($\sim 5$ times)
is the cancellation of the $\sigma$ and $\omega$ meson contributions
to $V_0$ (see eq. (\ref{depth}))
 which appears in the denominator of the relative variation
of $D$:  $\frac{\delta D}{D}=\frac{\delta V_0}{V_0} + ... $. 
However,  the $\sigma$ and $\omega$ mesons contribute with
 equal sign to the spin-orbit interaction constant $C$ \cite{C}. Therefore,
there is no ``cancellation'' enhancement here.
However, there is another efficient mechanism. The spin-orbit interaction
is inversely proportional to the nucleon mass $M_N$  squared,
 $C \propto 1/M_N^2$, $\frac{\delta C}{C}=-2 \frac{\delta M_N}{M_N}$. 
The nucleon mass depends on the quark masses:
$ \frac{\delta M_N}{M_N}=K_q \frac{\delta m_q}{m_q}+
K_s \frac{\delta m_s}{m_s}$ where $K_q$=0.045 and 
 $K_s$=0.19 in Refs. \cite{FS,FS2},$K_q$=0.037 and  $K_s$=0.011
 in Ref. \cite{thomas}, and $K_q$=0.064 in Ref. \cite{roberts}.
All three values of $K_q$ are close to the average value $K_q$=0.05.
However, different methods of calculations give very different values
of $K_s$. Fortunetely, this is not important since the strange
mass dependense of $\sigma$-meson is much stronger than that of proton
 (due to the SU(3) symmetry
 $\sigma$ contains valence $\bar s  s$
pair, another factor is strong repulsion of  $\sigma$  from the
 close  $K^+K^-,\bar K^0  K^0,\eta\eta $
 states \cite{FS}),
 also there is the ``cancellation'' enhancement of the $\sigma$ contribution 
to $D$. As a result we have
\begin{equation}\label{deltaDC}
\frac{\delta D}{D}- \frac{\delta C}{C}\approx 
\frac{\delta \alpha}{\alpha} + (0.11+0.10) \frac{\delta m_q}{m_q }
 -(3.5-0.2) {\delta m_s \over m_s}
\end{equation}  
 The final estimate for the relative variation of the $^{229}$Th
 transition frequency in eq. (\ref{delta1}) is
 \begin{equation}\label{deltaf}
\frac{\delta \omega}{\omega} \approx 10^5 (
4 \frac{\delta \alpha}{\alpha} +  \frac{\delta X_q}{X_q }
 -10 {\delta X_s \over X_s})\frac{3.5\,eV }{\omega}
\end{equation} 
where $X_q=m_q/\Lambda_{QCD}$ and $X_s=m_s/\Lambda_{QCD}$.
Thus we have here five to six orders enhancement in the relative
variation of the transition frequency. Another advantage is that the
 width of this nuclear transition is several orders of magnitude smaller
than a typical atomic clock width ($\sim$ Hz). 
Current atomic clocks limits on the variation of the fundamental constants
are approaching $10^{-15}$ per year. With these two enhancement factors
the result for  $^{229}$Th may be 7-10 orders of magnitude better.
We conclude
that this nuclear transition has an enormous potential for
a laboratory search for the variation of the fundamental constants.

This work is  supported by the Australian Research
Council.


\begin{thebibliography}{99}
\bibitem{comment} In principle, the shell effect may also produce
some small non-zero  deformation. Therefore, we should write
$\beta_2 =c_0 + c_1  \cdot \alpha + c_2  \cdot \alpha^2 ...$. However,
 $^{229}$Th is not so far from  the area of spherical nuclei
and the linear term must be important. This justifies our estimate.
\bibitem{comment1} Note that the suppression of the nucleon wave function
 at small separation due to the repulsive core reduces the depth of the effective potential $V$.
 However, this effect is not so important in the ratio ${\delta V \over V}$.
\end{thebibliography}
\end{document}